\documentclass[12pt]{article}

\usepackage{amsfonts}
\newcommand{\Nat}{\mathbb{N}}

\usepackage{graphicx}
\begin{document}

\title{Absence of barriers in dynamical triangulation}

\author{Bas V. de Bakker\thanks{email: bas@phys.uva.nl} \and Institute
  for Theoretical Physics, University of Amsterdam\\ Valckenierstraat
  65, 1018 XE Amsterdam, the Netherlands.}

\date{28 November 1994}

\maketitle

\renewcommand{\abstractname}{\normalsize Abstract}
\begin{abstract}
  \normalsize
  Due to the unrecognizability of certain manifolds there must exist
  pairs of triangulations of these manifolds that can only be reached
  from each other by going through an intermediate state that is very
  large.  This might reduce the reliability of dynamical
  triangulation, because there will be states that will not be reached
  in practice.  We investigate this problem numerically for the
  manifold $S^5$, which is known to be unrecognizable, but see no sign
  of these unreachable states.
\end{abstract}

{
\thispagestyle{myheadings}
\renewcommand{\thepage}{ITFA-94-35}

\clearpage
}

\section{Introduction}

Although the basis of it was laid in 1982 \cite{We82}, the dynamical
triangulation model of quantum gravity has in the last few years
received a lot of interest.  See e.g.\ \cite{AgMi92a,AmJu92}.  In this
model, the path integral over euclidean metrics is defined by a sum
over simplicial complexes.  This sum can then be approximated using
Monte Carlo techniques, where a computer program generates
appropriately weighted configurations of simplices.

To generate all these configurations we need an algorithm that is
ergodic, i.e.\ a set of moves that can transform any triangulation
into any other triangulation with the same topology.  A well known set
of moves that satisfy this condition are the so-called $(k,l)$ moves,
whose ergodicity was shown in \cite{GrVa92}.

\section{Noncomputability}

Unfortunately, the number of moves we need to get from one
configuration to another can be very large.  To be more precise, the
following theorem holds: if the manifold under consideration is
unrecognizable, then for any set of local moves the number of moves
needed to get from one configuration of $N$ simplices to another such
configuration is not bounded by a computable function of $N$.  This
was shown in ref.\ \cite{NaBe93}.  We will explain some of the terms
in this theorem in a way that is not mathematically precise, but
hopefully intuitively clear.  See \cite{NaBe93} for details.

A manifold is unrecognizable if, given a triangulation $A$ of this
manifold, there does not exist an algorithm that, given as input an
arbitrary triangulation $B$, can decide whether $A$ and $B$ are
homeomorphic.  The definition of unrecognizability is not important
for the rest of this article, it is only important to know that for
some manifolds the above theorem holds.  Certain four dimensional
manifolds are unrecognizable, but for the sphere $S^4$, which is
usually used in dynamical triangulation, this is not known.  It is
known, however, that the five dimensional sphere $S^5$ is
unrecognizable.

Local moves are moves that involve a number of simplices that is
bounded by a constant, in other words a number that does not grow with
the volume of the configuration.

A computable function is a function from $\Nat$ to $\Nat$ that can be
computed by a large enough computer.  Although the computable
functions are only an infinitesimally small fraction of all the
functions from $\Nat$ to $\Nat$, most functions one can think of are
computable.  A fast-growing example of a computable function would be
$N!!\cdots!$ with $N$ factorial signs.

The above theorem might seem a terrible obstacle for numerical
simulation, but the theorem says nothing about the behaviour of the
number of moves needed to generate any particular size of
configuration.  In fact, take any function $g(n)$ with the property
that it is not bounded by a computable function.  Replacing any finite
number of values of this function will result in another function
$g'(n)$ that is also not bounded by a computable function.

\section{Barriers}

\newcommand{\nint}{\ensuremath{N_{\mathrm{int}}(N)}}

From the theorem stated above it follows that for an unrecognizable
manifold the maximum size \nint\ of the intermediate configurations
needed to interpolate between any two configurations of size $N$ is
also not bounded by a computable function of $N$.  If \nint\ did have
such a bound, a bound on the number of possible configurations of size
less than or equal to \nint\ would be a bound on the number of moves
needed, which would violate the theorem.  A simple computable bound on
the number of configurations of size $N$ is $((d+1)N)!$, where $d$ is
the dimension of the simplices.

It was pointed out in \cite{AmJu94} that this means that for such a
manifold there must exist barriers of very high sizes between certain
points in configuration space.  Although the situation is not clear
from the theorem, it seems natural that these barriers occur at all
scales.  We can then apply the following method, which was formulated
in \cite{AmJu94}.  We start from an initial configuration with minimum
size.  For $S^4$ and $S^5$, there is a unique configuration of minimum
size with 6 and 7 simplices respectively.  We increase the volume to
some large number and let the system evolve for a while, which might
take it over a large barrier.  Next, we rapidly decrease the volume,
hoping to trap the configuration on the other side of this barrier.

We can check whether this has happened by trying to decrease the
volume even more.  If this brings us back to the initial
configuration, we have gone full circle and cannot have been trapped
at the other side of such a barrier.  Conversely, if we get stuck we
are apparently in a metastable state, i.e.\ at a point in
configuration space where the volume has a local minimum.

This was tried in \cite{AmJu94} for $S^4$, but no metastable states
were found.  To judge the significance of this, it is useful to
investigate the situation for a manifold which is known to be
unrecognizable.  It is rather difficult to construct a four
dimensional manifold for which this is known, but if we go to five
dimensions this is easy, because already the sphere $S^5$ is not
recognizable.

\section{Results}

Because my program for dynamical triangulation was written for any
dimension, it was not difficult to investigate $S^5$.  The description
by Catterall in \cite{Ca94} of his dynamical triangulation program for
arbitrary dimension turned out to be a very close description of mine,
presumably because both were based on ideas put forward in
\cite{BrMa93,Br93}.  The Regge-Einstein action in the five dimensional
model is
\begin{equation}
  S = \kappa_5 N_5 - \kappa_3 N_3
\end{equation}
where $N_i$ is the number of simplices of dimension $i$.  This is not
the most general action linear in $N_i$ in five dimensions as this
would take three parameters, but for the purposes of this paper this
is not relevant.

I generated 26, 24 and 8 configurations at $N_5 = 8000$, $16000$ and
$32000$ simplices respectively.  These were recorded each 1000 sweeps,
starting already after the first 1000 sweeps, were a sweep is $N_5$
accepted moves.  All configurations were made at curvature coupling
$\kappa_3 = 0$, making each configuration contribute equally to the
partition function, in other words making them appear equally likely
in the simulation.  Looking at the number of hinges $N_3$, the system
seemed to be thermalized after about 6000 sweeps, irrespective of the
volume.

The critical value of $\kappa_5$ (the bare cosmological constant)
below which the volume diverges was measured as explained in
\cite{CaKoRe94b,BaSm94b}.  It turned out to be 0.8252(4), 0.8366(5)
and 0.8446(8) for the three volumes used.  The last error cannot be
trusted, because of the low statistics at the largest volume.

Starting with these configurations, the volume was decreased by
setting $\kappa_5$ to a fixed number larger than the critical value.
We call this process cooling, because it attempts to reach a
configuration of minimum volume and thereby minimum action.

\begin{figure}[t]
\includegraphics[width=\textwidth]{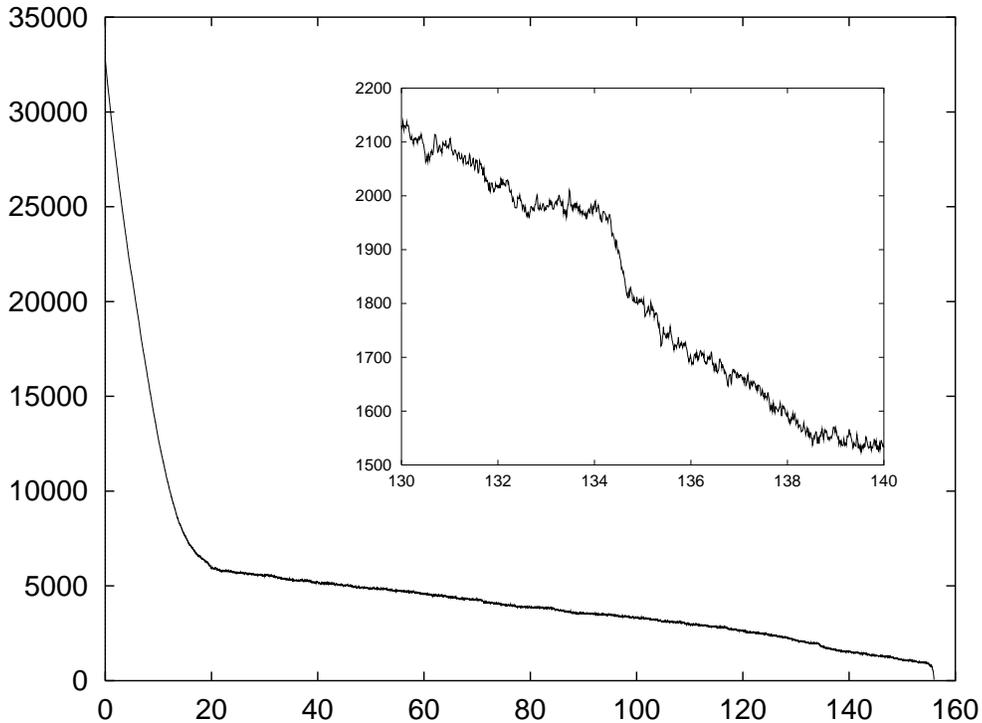}
\caption{A typical cooling run starting at $N_5 = 32000$, using
  $\kappa_5 = 2$.  The horizontal units are 1000 accepted moves.  The
  vertical axis is the number of 5-simplices.  The inset is a blowup
  of a small part of the curve.}
\label{run2fig}
\end{figure}
\begin{figure}[t]
\includegraphics[width=\textwidth]{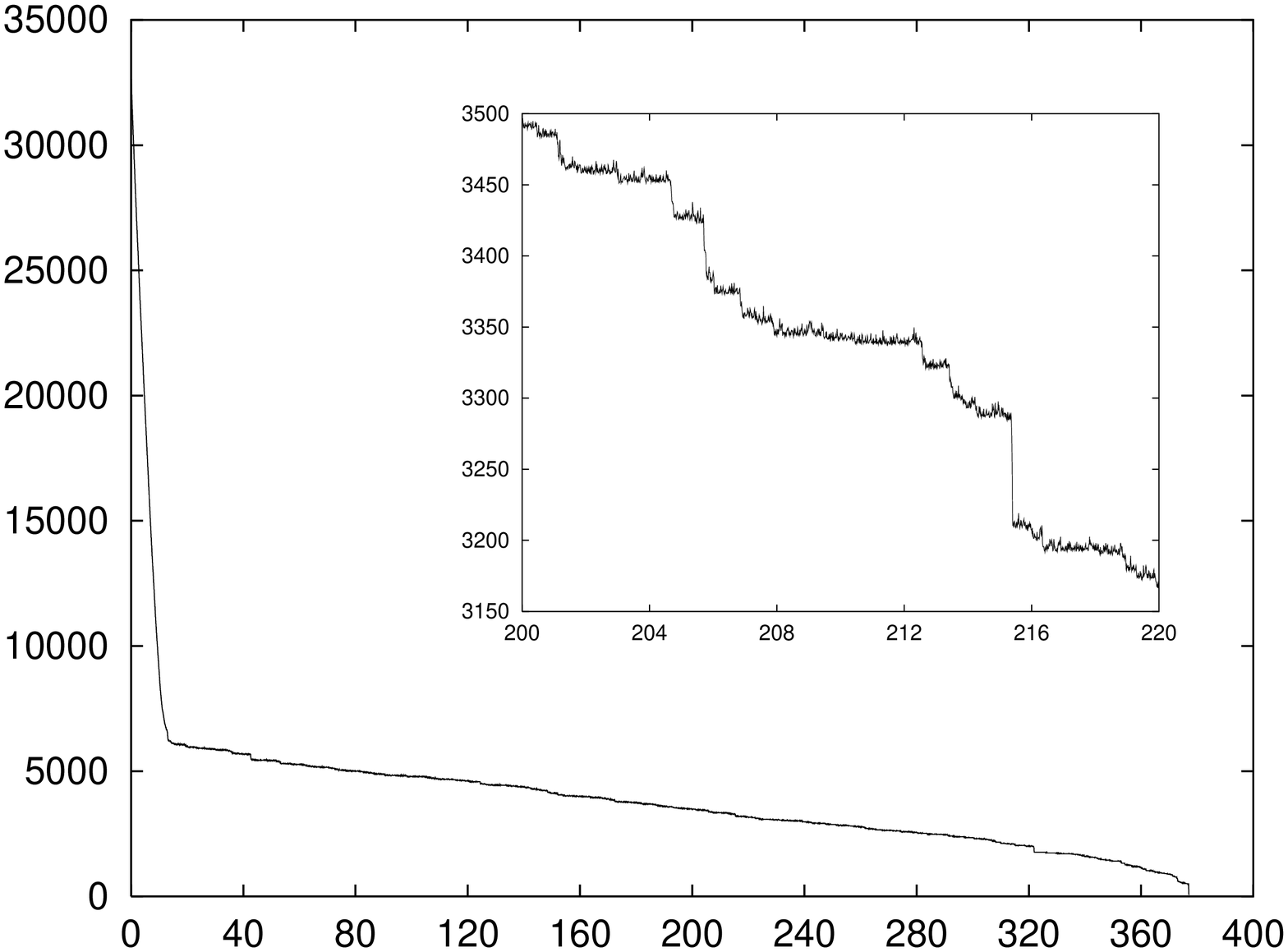}
\caption{As figure \protect\ref{run2fig}, but with $\kappa_5 =3$.}
\label{run3fig}
\end{figure}

For each configuration we cooled four times with $\kappa_5 = 2$ and
two times with $\kappa_5 = 3$.  For both values of $\kappa_5$ used one
of the runs is shown in figures \ref{run2fig} and \ref{run3fig}.  In
the insets we can see the typical volume fluctuations that occurred.
These were of the order 30 at $\kappa_5 = 2$ and 6 at $\kappa_5 = 3$.
The volume would first decrease very quickly until it reached roughly
a quarter of the starting value and then started to decrease much
slower.  In all cases the initial configuration of 7 simplices was
reached.  We also tried to use $\kappa_5 = 4$.  The same behaviour of
fast and slow cooling was seen, but the latter was so slow that due to
CPU constraints these had to be stopped before either a stable
situation or the minimal volume was reached.

There is an important difference between four and five dimensions.  In
four dimensions there is a move that leaves the volume constant.
Therefore the system can evolve at constant volume.  In five
dimensions this is not possible, because all moves change the volume.
In this case the volume has to fluctuate for the configuration to
change.  This is why much larger values of the cosmological constant
(such as were used in \cite{AmJu94}) would effectively freeze the
system.

\begin{figure}[t]
\includegraphics[width=\textwidth]{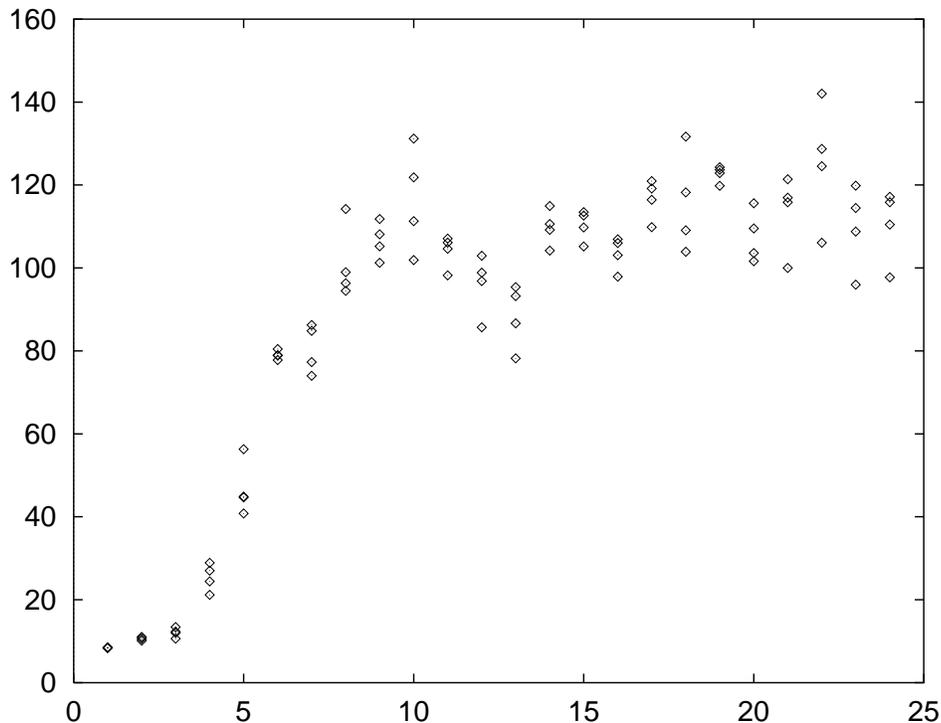}
\caption{Number of moves needed for cooling at $\kappa_5 = 2$ 
  as a function of the number of sweeps at the large volume of 16000
  simplices.  Horizontal units are 1000 sweeps, vertical units are
  1000 accepted moves.}
\label{timefig}
\end{figure}
Initially, before the system was thermalized, there was a strong
positive correlation between the time used to evolve the system at
large volume and the time needed to cool the system back to the
initial configuration.  The rates of fast and slow cooling did not
change, but the volume at which the slow cooling set in became larger.
Some of these times are shown in figure \ref{timefig} for the case of
16000 simplices.  This trend does not continue and the cooling times
seem to converge.

\section{Discussion}

So, contrary to expectation, no metastable states were seen.  Small
volume fluctuations were necessary, but these gave no indication of
the high barriers expected.

It is not clear why we don't see any metastable states.  There are
several possibilities.  First the barriers might be much larger than
32000 and we need to go to extremely large volumes before cooling.
Second, there might be no barriers much larger than the volume for the
volumes we looked at and the size of the intermediate configurations
needed still grows very slowly for the volumes considered, even though
this size is not bounded by a computable function.  Third, the
metastable regions in configuration space might be very small and the
chance that we see one is therefore also very small.

It was speculated in \cite{AmJu94} that the absence of visible
metastable states might indicate that $S^4$ is indeed recognizable.
The results shown in this paper for $S^5$ (which is known to be
unrecognizable) indicate that, unfortunately, the results for $S^4$
say nothing about its recognizability.  This, of course, in no way
invalidates the conclusion of \cite{AmJu94} that the number of
unreachable configurations of $S^4$ seems to be very small.

Recognizability is not the only thing that matters.  Even if $S^4$ is
recognizable, this says little about the actual number of $(k,l)$
moves needed to interpolate between configurations, except that this
number might be bounded by a computable function.

\end{document}